\documentstyle[12pt]{article}
\begin{document}
\title{\large \bf AN ANALOG OF THE VARIATIONAL DERIVATIVE AND CONSTRUCTIVE
NECESSARY INTEGRABILITY CONDITIONS FOR HYPERBOLIC EQUATIONS}
\author{S. Ya. Startsev\\
Mathematical Institute \\
Ufa center of the Russian Academy of Sciences\\
Chernyshevsky street 112, 450000 Ufa, Russia\\
e-mail: starts@imat.rb.ru}
\date{}
\maketitle
\vspace{5mm}

\begin{abstract}
An algorithm is constructed which allows to express conserved flows
of hyperbolic equations in terms of corresponding conserved densities and
to eliminate these flows from conservation laws of hyperbolic equations.
The application of this algorithm to canonical conservation laws gives
constructive necessary integrability conditions of hyperbolic equations in
terms of the generalized Laplace invariants of these equations.
\end{abstract}
\vspace{5mm}

Let us consider the hyperbolic equation
$$u_{xy}=F(x,y,u,u_{x},u_{y})\ . \eqno{(1)}$$

The partial derivatives ${\partial^{i+j} u}\over{\partial x^{i} \partial
y^{j}}$ with $i \cdot j \not=0$ can be eliminated by virtue of the equation
(1) and its differential consequences.  Therefore, we assume that all
functions depend on the variables $x,y,u,
u_{i} = {{\partial^{i} u}\over{\partial x^{i}}},
v_{i} = {{\partial^{i} u}\over{\partial y^{i}}} $.
We denote  the total derivatives with respect to $x$ and $y$ by virtue of
the equation (1) as $D_{x}$ and $D_{y}$, respectively. These total
derivatives are expressed in the variables $x,y,u,u_{i},v_{i}$ as
$$D_{x}={\partial\over{\partial x}}+u_{1}{\partial\over{\partial u}}+
\sum^{\infty }_{i=1}\left(u_{i+1}{\partial\over{\partial u_{i}}}
+D^{i-1}_{y}(F){\partial\over{\partial v_{i}}}\right) \ ,$$
$$D_{y}={\partial\over{\partial y}}+v_{1}{\partial\over{\partial u}}+
\sum^{\infty }_{i=1}\left(v_{i+1}{\partial\over{\partial v_{i}}}
+D^{i-1}_{x}(F){\partial\over{\partial u_{i}}}\right) \ . $$

Let us consider the differential operator
$$ M= D_{x} D_{y} + a D_{x} + b D_{y} + c\ , \eqno{(2)}$$
where $a,b$ and $c$ are functions.

\medskip
{\bf Definition 1.\ }{\sl The functions
$$K_{0}=D_{y}(b)+ a b - c, \enskip H_{0}=D_{x}(a)+ a b - c$$
are called the main Laplace x-invariant and y-invariant of the operator
(2), respectively.

The subsequent Laplace y-invariants $H_{i}$ are defined by the formula
$$H_{i+1}=2H_{i}-D_{x}(S_{i})-H_{i-1}\ ,$$
where $H_{-1}=K_{0}$ and $S_{i}$ is a solution of the equation $H_{i} S_{i}
=D_{y}(H_{i})$.

The subsequent Laplace x-invariants $K_{i}$ are defined by the formula
$$K_{i+1}=2K_{i}-D_{y}(R_{i})-K_{i-1}\ ,$$
where $K_{-1}=H_{0}$ and $R_{i}$ is a solution of the equation $K_{i} R_{i}
=D_{x}(K_{i})$.

The Laplace invariants of the linearization operator
$$ L= D_{x} D_{y} - F_{u_{x}} D_{x} - F_{u_{y}} D_{y} - F_{u} \eqno{(3)}$$
of the equation (1) are called the Laplace invariants of this equation.}
\medskip

Let us denote
$$\begin{array}{ll}
a_{0}=a,\  a_{i+1}=a_{i}-S_{i};& b_{0}=b,\  b_{i+1}=b_{i}-R_{i}; \\
M_{i}=(D_{y}+a)(D_{x}+b_{i})-K_{i};& M_{-i}=(D_{x}+b)(D_{y}+a_{i})-H_{i}; \\
\nabla_{i}=D_{x}+b_{i};&\Delta_{i}=D_{y}+a_{i}; \\
A_{-1}=1,& A_{i}=\Delta_{i} \circ A_{i-1}; \\
B_{-1}=1,& B_{i}=\nabla_{i} \circ B_{i-1}; \\
C_{0}=1,& C_{i+1}=\Delta_{i+1} \circ C_{i}, \
\end{array}$$
where $i \ge 0$.

It is easy to prove the relationship
$$C_{i} \circ M = \nabla_{0} \circ A_{i}-î_{i} A_{i-1}\ ,\ i \ge 0
\eqno {(4)} $$
by induction on $i$.

\medskip
{\bf Definition 2.\ }{\sl  A function $f$ is called a symmetry of the
equation (1) if $L(f)=0$, where $L$ is defined by the formula (3).}
\medskip

Let $f$ be a symmetry of the equation (1) and
$$p= D_{x}(\gamma )\ ,\quad p_{u_{i}} = 0,\ i>n\ .   \eqno{(5)} $$
It is easy to verify that the vector field
$$\partial_{f}=f{\partial\over{\partial u}}+\sum^{\infty }_{i=1} \left(
D^{i}_{y}(f){\partial\over{\partial v_{i}}}+ D^{i}_{x}(f)
{\partial\over{\partial u_{i}}}\right)$$
commutes with the  total derivatives $D_{x}$ and $D_{y}$. Therefore,
applying $\partial_{f}$ to (5) we obtain that the operator
$p_{*} - D_{x} \circ \gamma _{*}$, where $q_{*}$ denotes the linearization
operator of the function $q$,
$$q_{*}={{\partial q}\over{\partial u}}+\sum^{\infty}_{i=1}\left( {{\partial
q}\over{\partial v_{i}}}D^{i}_{y}+ {{\partial Q}\over{\partial
u_{i}}}D^{i}_{x}\right)\ \ ,$$
maps any symmetry into zero. Hence,
$$ p_{*} - D_{x} \circ \gamma _{*} = 0 \quad mod \quad
\{L,D_{x} \circ L, D_{y} \circ L,\dots \} \ . \eqno{(6)} $$

We rewrite
$$\begin{array}{ll}
p_{*}&=\xi _{0}+ \sum^{\infty}_{i=1} (\xi _{-i} A_{i-1} + \xi _{i}
B_{i-1}) \ , \\
\gamma_{*}&=\eta _{0} +\sum^{\infty}_{i=1} (\eta _{-i} A_{i-1} +
\eta _{i} B_{i-1}) \ . \
\end{array} \eqno{(7)}$$
Using the formula (4) we rewrite (6) as
$$\begin{array}{ll}
\xi _{n} &= \eta_{n-1} \ , \\
\xi _{i} &= (D_{x}-b_{i}) (\eta _{i}) +\eta _{i-1},\ i=\overline{1,n-1}
\ , \\
\xi _{-i} &=(D_{x}+F_{u_{y}}) (\eta _{-i}) + \eta _{-(i+1)} H_{i},\
i \ge 0 \ , \
\end{array} \eqno{(8)}$$
where $H_{i}$ are the Laplace y-invariants of (1),
$b_{i} = - F_{u_{y}} - \sum^{i-1}_{j=0} R_{j}$.

Let us denote
$$\begin{array}{cl}
\delta _{n-1} (p)& = \xi _{n} \ , \\
\delta _{i} (p) &= \xi _{i+1} - (D_{x}-b_{i+1}) (\delta _{i+1} (p)),\
i=\overline{-1,n-2} \ , \\
\delta _{-i} (p) &= H_{i-2} \dots H_{0} \xi _{1-i} -
(D_{x}-{\hat{b}}_{i-1}) (\delta _{1-i} (p)),\ i \ge 2 \ , \
\end{array} \eqno{(9)}$$
where ${\hat{b}}_{i} = -F_{u_{y}} + \sum^{i-1}_{j=0} Q_{j},\
D_{x}(H_{j}) = Q_{j} H_{j} $. The equations (8) imply in this notation
that
$$\begin{array}{cl}
\eta _{i} &= \delta _{i} (p),\quad i=\overline{0,n-1}, \\
\eta _{-i} H_{i-1} \dots H_{0} &= \delta _{-i} (p),\quad i>0 \ . \
\end{array}$$
Since there exist $k$ such that $\gamma _{v_{i}} = 0$ for all $i>k$ and,
therefore, $\eta_{-i} = 0$, \hbox{$i>k$}, we prove the following

\medskip
{\bf Proposition 1.\ }{\sl If $p \in Im\ D_{x}$, then there exist $k$ such
that $\delta _{-i} (p) = 0$ for all $i>k$.}
\medskip

Let us consider a conservation law
$$ D_{y}(p) = D_{x}(\gamma ),\quad p_{u_{i}}=0,\  i>n \ .$$
Repeating the slightly modified above calculation we obtain
$$\begin{array}{rl}
(D_{y}+F_{u_{x}}) (\xi _{n}) &= \eta _{n-1}\ , \\
(D_{y}+F_{u_{x}}) (\xi _{i}) + \xi _{i+1} K_{i} &=
(D_{x} - b_{i}) (\eta _{i}) + \eta _{i-1}\ ,\ i=\overline{1,n-1}, \\
(D_{y}+F_{u_{x}}) (\xi _{0}) + \xi _{1} K_{0} &=
(D_{x} + F_{u_{y}}) (\eta _{0}) + \eta _{-1} H_{0}\ , \\
(D_{y}-a_{i}) (\xi _{0}) + \xi _{1-i} &=
(D_{x} + F_{u_{y}}) (\eta _{-i}) + \eta _{-(i+1)} H_{i}\ ,\ i \ge 1 \ , \
\end{array}\eqno{(10)}$$
where $\xi _{i}$ and $\eta _{i}$ are defined by (7), $K_{i}$ are the
Laplace invariants of the equation (1), $a_{i} = - F_{u_{x}} -
\sum_{j=0}^{i-1} S_{j}$. Taking into account
$$\begin{array}{rl}
(D_{x} - b_{i-1}) \circ {(L_{i})}^{\top } & = {(L_{i-1})}^{\top } \circ
(D_{x} - b_{i})\ , \\
H_{i-1} \dots H_{0} (D_{y} - a_{i}) & = (D_{y} + F_{u_{x}}) \circ
H_{i-1} \dots H_{0}\ , \\
H_{i-1} \dots H_{0} (D_{x} + F_{u_{y}}) & = (D_{x} - \hat{b} _{i}) \circ
H_{i-1} \dots H_{0}\ , \\
(D_{x} - \hat{b} _{i}) \circ {(L^{\top})}_{i-1} & = {(L^{\top})}_{i} \circ
(D_{x} - \hat{b} _{i-1})\ , \
\end{array}$$
where $i \ge 1$, $M^{\top}$ denotes the formally adjoint to
$M$ operator
$$ M^{\top} = D_{x} D_{y} - a D_{x} - b D_{x} + c - D_{x}(a) - D_{y}(b)\ , $$
we deduce from (10)
$$\begin{array}{cl}
\eta _{n-1} & = (D_{y}+F_{u_{x}}) (\delta _{n-1} (p))\ , \\
\eta _{i} & = (D_{y}+F_{u_{x}}) (\xi _{i+1}) - {(L_{i+1})}^{\top}
(\delta _{i+1} (p))\ ,\ i=\overline{0,n-2}\ , \\
\eta _{-1} H_{0} & = (D_{y}+F_{u_{x}}) (\xi _{0}) - L^{\top}
(\delta _{0} (p))\ , \\
\eta _{-i} H_{i-1} \dots H_{0} & =(D_{y}+F_{u_{x}}) ( H_{i-2} \dots H_{0}
\xi _{1-i}) - {(L^{\top})}_{i-1} (\delta _{1-i} (p))\ ,\ i \ge 2 \ , \
\end{array}$$
where $\delta _{i} (p)$ are defined by (9).

Thus, we prove

\medskip
{\bf Proposition 2.\ }{\sl If $D_{y} (p) \in Im\ D_{x}$, then there
exist $k$ such that \hbox{${(L^{\top})}_{i} (\delta _{-i} (p)) = 0$}
for all $i>k$.}
\medskip

{\bf Proposition 3.\ }{\sl Let the equation (1) admits a symmetry $f$,
$f_{v_{k}} \ne 0$, $f_{v_{i}} = 0$ for all $i > k$. Then
$$\begin{array}{rll}
D_{x}(f_{v_{k}}) & = 0 & \hbox{if\ } k \ge 1\ ; \\
D_{y}(F_{u_{y}}) & = D_{x} \left( {{f_{v_{k-1}}}\over{k f_{v_{k}}}}\right)
& \hbox{if\ } k \ge 3 \qquad (11)\ ; \\
D_{y}(\root k \of {f_{v_{k}}} (F_{u} + F_{u_{x}} F_{u_{y}})) &
\in Im\ D_{x} & \hbox{if\ } k \ge 5 \ .  \
\end{array}$$}
\medskip

Differentiating the relationship $L(f) = 0$ with respect to $v_{k+1},
v_{k}$ and $v_{k-1}$ we easily prove this proposition.

It is easy to verify by induction on $i$ that
$$ \delta _{-i} (F_{u_{y}}) = h_{i-1} H_{i-2} \dots H_{0}\ , $$
where $h_{i-1} ={ \left( H_{i-1} \right)}_{v_{i}}$, $ i \ge 1$.

The application of proposition 2 to the canonical conservation low (11) and
the formula ${(L^{\top})}_{i} \circ H_{i-1} \dots H_{0} = H_{i-1} \dots H_{0}
{(L_{-i})}^{\top}$ give

\medskip
{\bf Theorem.\ }{\sl If the equation (1) has a symmetry $f$, $f_{v_{k}}
\ne 0$, $f_{v_{i}} = 0$ for all $i > k$, $k \ge 3$, then
$${(L_{-k})}^{\top} \left( {{h_{k-1}} \over {H_{k-1}}} \right) = 0 $$
\centerline{or}
\centerline{$H_{i} = 0$ for some $i$, $0 \le i \le k-1 $.}}
\medskip

Moreover, we can express $\gamma _{u_{i}}$ and $\gamma _{v_{i}}$ in terms
of $\eta _{i}$ and obtain a first order partial differential system on
$\gamma $. The compatibility conditions of this system corresponding to a
canonical conservation law give additional integrability conditions of the
equation (1).

\end{document}